\documentclass[12pt]{article}

\usepackage{jheppub,amsbsy,amssymb,latexsym,amsfonts,amsmath,epsfig,multicol,bbm}

\allowdisplaybreaks

\title{The classical origin of quantum affine algebra in squashed sigma models}

\author{Io Kawaguchi$^{\ast}$\footnote{E-mail:~io@gauge.scphys.kyoto-u.ac.jp}, 
Takuya Matsumoto$^{\dagger,\ddagger}$\footnote{E-mail:~tmatsumoto@usyd.edu.au} 
and Kentaroh Yoshida$^{\ast}$\footnote{E-mail:~kyoshida@gauge.scphys.kyoto-u.ac.jp}}
\affiliation{$^{\ast}$Department of Physics, Kyoto University, Kyoto 606-8502, Japan \\
$^{\dagger}$School of Mathematics and Statistics, University of Sydney, NSW 2006, Australia \\
$^{\ddagger}$Graduate School of Mathematics, Nagoya University, Nagoya 464-8602, Japan}

\arxivnumber{1201.3058[hep-th]}

\abstract{
We consider a quantum affine algebra realized in two-dimensional non-linear sigma models 
with target space three-dimensional squashed sphere. 
Its affine generators are explicitly constructed and the Poisson brackets are computed. 
The defining relations of quantum affine algebra in the sense of 
the Drinfeld first realization are satisfied at classical level. 
The relation to the Drinfeld second realization is also discussed including higher conserved charges. 
Finally we comment on a semiclassical limit of quantum affine algebra at quantum level. 
}


\begin{document}

\maketitle

\section{Introduction}

Integrable quantum field theories are wonderful laboratories to develop non-perturbative methods 
such as strong/weak-coupling dualities. 
Some examples are principal chiral models and $O(N)$-invariant non-linear sigma models, 
and these are integrable at both classical and quantum mechanical level \cite{Luscher1,Luscher2}. 
The integrability is closely related to the symmetric coset structure of target space \cite{AAR}. 

\medskip 

For symmetric coset target spaces, a general prescription to construct an infinite number of 
classical conserved charges is well known \cite{Luscher2,AAR,BIZZ} and 
the charges obtained along it form the Yangian algebra \cite{Bernard-Yangian,MacKay}, 
mathematically formulated by Drinfeld \cite{Drinfeld}. Then  
the quantum integrability can be argued by checking whether the conservation laws 
of the charges are anomalous or not \cite{GW,AFG}. When the system is quantum mechanically integrable, 
the Bethe ansatz technique works well as in principal chiral models \cite{PW}.

\medskip 

The integrability plays an important role in the recent study of AdS/CFT \cite{Maldacena} 
(For a comprehensive review see \cite{review}). In particular, 
the classical integrability of sigma model on the string theory side 
is discussed in \cite{BPR} and it is inherited from the behind structure of AdS/CFT, the parent integrable spin chain. 
It would be a nice direction to seek for a generalization of AdS/CFT preserving the sigma model integrability. 
Within a class of symmetric spaces including the supersymmetric extension, 
it has been done quite generally in \cite{Zarembo}. 
Hence the next is to consider non-symmetric cosets as a generalization. 

\medskip 

There are some integrable deformations of AdS/CFT, for example, 
$\beta$-deformation \cite{LS,LM,Frolov,Berenstein} and $q$-deformation of 
the world-sheet S-matrix \cite{BK, BGM, Hollowood}. 
However, we are interested in another class of integrable deformations concerning 
warped AdS spaces and squashed spheres in three dimensions, which are represented by non-symmetric cosets.  
Since warped AdS spaces are obtained via double Wick rotations from squashed spheres\footnote{
The Wick rotations are detailed in \cite{KY}. The Poisson structure is also obviously preserved 
and the non-compactness is not relevant to the classical analysis discussed here.},  
we are confined to squashed spheres here. The main subject in this paper is to gain more insight into 
the classical integrable structure of two-dimensional non-linear sigma models with squashed spheres as target space\footnote{  
The sigma models are often referred here to as ``squashed sigma models'' as an abbreviation.}.  

\medskip 

Although the squashed sigma models are well known as an integrable model of the trigonometric class, 
it has been shown that a Yangian symmetry $Y(sl(2))$ is realized even after the squashing in a series of works 
\cite{KY,KOY,KYhybrid} (For a short summary see \cite{KY-summary}). 
This result may sound curious. However, in fact, there exist two descriptions to describe the classical dynamics, 
i) the trigonometric description and ii)  the rational description. 
That is, it is possible to construct two types of Lax pair, both of which lead to the identical classical equations of motion. 
The two descriptions are related to each other via a non-local map.  
Furthermore, a finite-dimensional quantum group symmetry $U_q(sl(2))$ \cite{Drinfeld,Jimbo} 
is realized at classical level in terms of non-local currents corresponding to the broken generators 
due to the squashing of target space. 
This is nothing but a classical origin of quantum group symmetry \cite{Drinfeld,Jimbo}. 
The explicit relation between the algebraic deformation parameter $q$ of the quantum group $U_q(sl(2))$ 
and the geometric deformation parameter of squashed three sphere is also given by \cite{KYhybrid}. 

\medskip 

In this paper, as a generalization of the previous result, it is shown that the quantum group symmetry $U_q(sl(2))$ 
is enhanced to the quantum affine algebra $U_q(\widehat{sl(2)})$ at classical level. An infinite number of classical conserved charges 
are derived by expanding the monodromy matrix constructed with the Lax pair in the trigonometric description \cite{FR}. 
The infinite tower structure of the charges with the Poisson bracket 
turns out to be a classical analogue of quantum affine algebra $U_q(\widehat{sl(2)})$ \cite{Drinfeld}. 
Firstly, we show that the Poisson brackets of the level $0,\pm1$ charges satisfy the defining relations of the 
quantum affine algebra $U_q(\widehat{sl(2)})$ in the sense of the Drinfeld first realization. 
Secondly, we argue that including all of the higher charges recast the tower structure 
into the Drinfeld second realization. 

\medskip 

This paper is organized as follows. In section 2 we introduce the classical action of squashed sigma models and 
the monodromy matrix in the trigonometric description. 
In section 3 an infinite set of conserved (non-local) charges are derived by expanding the monodromy matrix. 
In section 4, by evaluating the classical Poisson brackets, 
we show that they actually coincide with the defining relations of quantum affine algebra $U_q(\widehat{sl(2)})$
in the sense of the Drinfeld first realization. 
Then we argue the relation between higher conserved charges and the Drinfeld second realization at classical level. 
In section 5 we comment on a semiclassical limit of quantum affine algebra realized at quantum level. 
Section 6 is devoted to conclusion and discussion. 
In Appendix A we give a proof of the classical $q$-Serre relations, 
which are a part of the defining relations in the Drinfeld first realization 
and impose constraints on the higher conserved charges.

\section{Squashed sigma model and monodromy matrix}

We consider two-dimensional non-linear sigma models with target space squashed sphere in three dimensions.
The classical action is given by 
\begin{eqnarray}
S[J] &=&\!\! \int\!\!\!\!\int\!\!dtdx\Bigl[
{\rm Tr}\left(J_{\mu} J^{\mu}\right)
-2C\,{\rm Tr}\!\left(T^{3}J_{\mu}\right)\!{\rm Tr}\!\left(T^{3}J^{\mu}\right)
\Bigr]\,, \nonumber \\
&&J_{\mu} \equiv g^{-1}\partial_{\mu}g\,,\qquad g\in SU(2)\,. 
\label{sigma}
\end{eqnarray}
The coordinates and metric of base space are $x^{\mu}=(t,x)$ and $\eta_{\mu\nu}={\rm diag}(-1,+1)$\,. 
Suppose that the value of $C\in\mathbb{R}$ is restricted to $C > -1$ 
so that the sign of kinetic term is not flipped. 
The $SU(2)$ Lie algebra generators $T^a~(a=1,2,3)$ satisfy  
\begin{eqnarray}
[T^a,T^b] = \varepsilon^{ab}_{~~c}T^c\,, \qquad {\rm Tr}(T^aT^b) = - \frac{1}{2}\delta^{ab}\,,
\end{eqnarray}
where $\varepsilon^{ab}_{~~c}$ is the totally anti-symmetric tensor. 

\medskip 

This system has the $SU(2)_{\rm L}\times U(1)_{\rm R}$ symmetry. The non-zero value of $C$ breaks 
the original $SU(2)_{\rm L} \times SU(2)_{\rm R}$ symmetry of round $S^3$ to $SU(2)_{\rm L}\times U(1)_{\rm R}$\,. 
As a matter of course, the $SU(2)_{\rm L} \times SU(2)_{\rm R}$ symmetry recovers when $C =0$\,. 

\medskip 

Note that the Virasoro and periodic boundary conditions are not imposed here, though we have some applications 
in string theory in our mind.  Instead of them, we impose the boundary condition that the group variable $g(x)$ 
approaches a constant element rapidly as it goes to spatial infinity like 
\begin{eqnarray}
g(x) \to g_{\infty}~:~\mbox{const.} \qquad (x \to \pm \infty)\,. \label{bc}
\end{eqnarray}
That is, $J_{\mu}(x)$ vanishes rapidly as $x\to \pm \infty$\,. 

\medskip 

The classical equations of motion are obtained in the usual manner as 
\begin{eqnarray}
\partial^{\mu}J_{\mu} - 2C {\rm Tr}(T^3\partial^{\mu}J_{\mu})T^3 - 2C\,{\rm Tr}(T^3J_{\mu})[J^{\mu},T^3] =0\,.
\label{eom}
\end{eqnarray}
It is possible to construct two types of Lax pair which lead to the identical equations of motion (\ref{eom}) 
and these are equivalent through a non-local map \cite{KYhybrid}. 
That is, there are two equivalent ways in describing the classical dynamics, 
1) the rational description based on $SU(2)_{\rm L}$ and 2) the trigonometric description based on $U(1)_{\rm R}$\,. 

\medskip 

We work below in the trigonometric description, 
where the Lax pair is given by \cite{FR} 
\begin{eqnarray}
L^{R}_t(x;\lambda) &=& \sum_{a=1}^3\left[ 
w_a(\lambda + \alpha) S^a - w_a(\lambda - \alpha) \bar{S}^a 
\right] T^a\,, \nonumber \\ 
L^{R}_x(x;\lambda) &=& \sum_{a=1}^3\left[ 
w_a(\lambda + \alpha) S^a + w_a(\lambda - \alpha) \bar{S}^a 
\right] T^a\,.  \label{Lax-R} 
\end{eqnarray}
Here $\lambda\in\mathbb{C}$ is a spectral parameter and $w_a(\lambda)$ is defined as 
\begin{eqnarray}
w_1(\lambda) = w_2(\lambda) \equiv \frac{1}{\sinh\lambda}\,, \qquad w_3(\lambda) \equiv \coth\lambda\,. 
\end{eqnarray}
Then $S^{a}$ and $\bar{S}^{a}$ are related to $J_{\mu}^a \equiv -2 {\rm Tr}(T^a J_{\mu})$ like 
\begin{eqnarray}
&& J_t^3 = (w_1(2\alpha)+w_3(2\alpha))(S^3 + \bar{S}^3)\,, \nonumber \\
&& J_x^3 = (w_1(2\alpha)+w_3(2\alpha))(S^3 - \bar{S}^3)\,,  \nonumber \\ 
&& J_t^{1,2} = \sqrt{2 w_1(2\alpha)(w_1(2\alpha)+w_3(2\alpha))}\,(S^{1,2} + \bar{S}^{1,2})\,, \nonumber \\ 
&& J_x^{1,2} = \sqrt{2 w_1(2\alpha)(w_1(2\alpha)+w_3(2\alpha))}\,(S^{1,2} - \bar{S}^{1,2})\,, \nonumber 
\end{eqnarray}
and the parameter $\alpha$ is written in terms of the squashing parameter $C$ as  
\begin{eqnarray}
\cosh\alpha \equiv \frac{1}{\sqrt{1+C}}\,,\qquad
\sinh\alpha \equiv \frac{i\sqrt{C}}{\sqrt{1+C}}\,. \nonumber 
\end{eqnarray}
Note that $\alpha$ is pure imaginary for $C>0$ and real for $C<0$\,. When $C=0$\,, $\alpha=0$\,. 

\medskip 

The commutator 
\[
[\partial_t + L_t^R(x;\lambda),\partial_x + L_x^R(x;\lambda)] =0
\]
leads to the equations of motion in (\ref{eom}) as well as the Maurer-Cartan equation 
\[
dJ + J \wedge J = 0\,. 
\]
Then the monodromy matrix $U^R(\lambda)$ is defined as
\begin{eqnarray}
U^R(\lambda)&\equiv&{\rm P}\exp\left[-\int^{\infty}_{-\infty}\!\!\!dx\,L^R_x(x;\lambda)\right]\,.  
\label{monodromy}
\end{eqnarray}
It is straightforward to show that this is a conserved quantity,   
\[
\frac{d}{dt}U^R(\lambda) =0\,. 
\]
It would be helpful later to use the following form of $L^R_x(x;\lambda)$\,, 
\begin{eqnarray}
L^R_x(x;\lambda)&=&\frac{\sinh\alpha}{\sinh(\lambda-\alpha)\sinh(\lambda+\alpha)}
\biggl[T_3\left(\frac{\sinh\lambda\cosh\lambda}{\cosh\alpha}J^3_t(x)-\sinh\alpha J^3_x(x)\right) \nonumber \\
&& + \left.T_+\left(\cosh\alpha\sinh\lambda J^+_t(x)-\sinh\alpha\cosh\lambda J^+_x(x)\right)\right. \nonumber \\
&& + T_-\left(\cosh\alpha\sinh\lambda J^-_t(x)-\sinh\alpha\cosh\lambda J^-_x(x)\right)\biggr]\,, \nonumber
\end{eqnarray}
where we have introduced the following notations,
\[
T^{\pm} \equiv \frac{1}{\sqrt{2}}(T^1 \pm i T^2) = T_{\mp}\,. 
\]
An infinite number of conserved charges are obtained by expanding the monodromy matrix $U^R(\lambda)$ with respect to $\lambda$ 
around an expansion point. The expression of charges depends on expansion points.

\section{Expanding monodromy matrix}

Let us expand the monodromy matrix $U^R(\lambda)$ with the complex parameter $z=e^{-\lambda}$. 
Depending on the regions of the complex plane, we obtain the following two expansions 
\begin{align}
\mbox{i)} \qquad 
U^R(\lambda)&=e^{\bar{u}_{0}}\exp\left[\sum_{n=1}^\infty z^{n}\bar{u}_{n}\right]  \qquad {\rm for} \quad|z|<1
\quad{\rm (or}\quad {\rm Re} (\lambda)>0)\\
\mbox{ii)} \qquad U^R(\lambda)&=e^{u_{0}}\exp\left[\sum_{n=1}^\infty z^{-n}u_{n}\right]  \qquad {\rm for} \quad|z|>1
\quad{\rm (or}\quad {\rm Re} (\lambda)<0)
\end{align}
Corresponding the expanding coefficients $\bar{u}_{0}$, $\bar{u}_{n}$ and $u_{0}$, $u_{n}$ ($n\geq 1$), 
we would define the conserved charges $\bar{Q}^{R,a}_{(n)}$ and $Q^{R,a}_{(n)}$ respectively, where superscript $a$ 
runs $\pm,3$ and denotes the triplet generators of $sl(2)$\,. 
An infinite number of conserved non-local charges are obtained systematically at classical level. 
These charges are nothing but the generators of quantum affine algebra $U_q(\widehat{sl(2)})$ as we will discuss later. 

\subsection{Expansion i)}

Let us consider the expansion i). 
Then the spatial component of the Lax pair is expanded around $z=0$ like 
\begin{eqnarray}
L^R_x(x;\lambda)&=&i\sqrt{C}T_3J^3_t(x) \nonumber \\
&& + z\left[T_+\left(\frac{2i\sqrt{C}}{1+C}J^+_t(x)+\frac{2C}{1+C}J^+_x(x)\right)
\right. \nonumber \\ && \left.
\qquad + T_-\left(\frac{2i\sqrt{C}}{1+C}J^-_t(x)+\frac{2C}{1+C}J^-_x(x)\right)\right]\nonumber \\
&& + z^2T_3\left(\frac{2i\sqrt{C}(1-C)}{1+C}J^3_t(x)+\frac{4C}{1+C}J^3_x(x)\right) \nonumber \\
&& + z^3\left[T_+\left(\frac{2i\sqrt{C}(1-3C)}{(1+C)^2}J^+_t(x)-\frac{2i\sqrt{C}(3-C)}{(1+C)^2}J^+_x(x)\right)\right. \nonumber \\
&& \qquad + \left.T_-\left(\frac{2i\sqrt{C}(1-3C)}{(1+C)^2}J^-_t(x)-\frac{2i\sqrt{C}(3-C)}{(1+C)^2}J^-_x(x)\right)\right] 
+ {\mathcal O}(z^4)\,. \nonumber 
\end{eqnarray}
The expanded monodromy matrix is 
\begin{eqnarray}
U^R(\lambda)&=&{\rm P}\exp\left[-\int^{\infty}_{-\infty}\!\!\!dxL^R_x(x;\lambda)\right] \nonumber \\
&=& {\rm e}^{\bar{u}_0} \left[1 + z \bar{u}_1 + z^2\left(\bar{u}_2 + \frac{1}{2}(\bar{u}_1)^2\right) 
\right. \nonumber \\ && \left.  
+z^3\left(\bar{u}_3+\frac{1}{2}(\bar{u}_2 \bar{u}_1 + \bar{u}_1 \bar{u}_2)+\frac{1}{6}(\bar{u}_1)^3\right)+{\mathcal O}(z^4)\right]\,, \nonumber 
\end{eqnarray}
where $\bar{u}_i~(i=0,1,2,3,\ldots)$ are 
\begin{eqnarray}
&& \bar{u}_0 = i\gamma T_3 \bar{Q}^{R,3}_{(0)}\,, \qquad 
\bar{u}_1 = -2i\gamma\left(T_-{\rm e}^{-\gamma \bar{Q}^{R,3}_{(0)}/2} Q^{R,-}_{(1)} 
+ T_+{\rm e}^{\gamma \bar{Q}^{R,3}_{(0)}/2} \widetilde{Q}^{R,+}_{(1)}\right)\,, \nonumber \\ 
&& \bar{u}_2 = 2i\gamma^2T_3 \bar{Q}^{R,3}_{(2)}\,, \qquad 
\bar{u}_3 = 2i\gamma^3\left(T_-{\rm e}^{-\gamma \bar{Q}^{R,3}_{(0)}/2} Q^{R,-}_{(3)} 
+ T_+ {\rm e}^{\gamma \bar{Q}^{R,3}_{(0)}/2} \widetilde{Q}^{R,+}_{(3)}\right)\,, \nonumber 
\end{eqnarray}
and a new parameter $\gamma$ is defined in terms of $C$ as 
\begin{eqnarray}
\gamma \equiv \frac{\sqrt{C}}{1+C}\,.
\end{eqnarray}
The conserved charges obtained up to the fourth order of $z$ 
are summarized below:   
\begin{eqnarray}
\bar{Q}^{R,3}_{(0)} &=& -\int^{\infty}_{-\infty}\!\!\!dx\,j^{R,3}_t(x)\,, \nonumber \\
Q^{R,-}_{(1)} &=& \int^{\infty}_{-\infty}\!\!\!dx\,j^{R,-}_t(x)\,, 
\qquad 
\widetilde{Q}^{R,+}_{(1)} = \int^{\infty}_{-\infty}\!\!\!dx~\widetilde{j}^{R,+}_t(x)\,, \nonumber \\
\bar{Q}^{R,3}_{(2)} &=& \int^{\infty}_{-\infty}\!\!\!dx\int^{\infty}_{-\infty}\!\!\!dy~\epsilon(x-y)j^{R,-}_t(x)
\widetilde{j}^{R,+}_t(y) 
+2i\int^{\infty}_{-\infty}\!\!\!dxj^{R,3}_x(x) + \frac{1-C}{\sqrt{C}}\bar{Q}^{R,3}_{(0)}\,, \nonumber \\
Q^{R,-}_{(3)} &=& \frac{1}{2} \int^{\infty}_{-\infty}\!\!\!dx\int^{\infty}_{-\infty}\!\!\!dy
\int^{\infty}_{-\infty}\!\!\!dz\, \epsilon(x-y)\epsilon(x-z)\widetilde{j}^{R,+}_t(x)j^{R,-}_t(y)j^{R,-}_t(z) 
\nonumber \\ &&
-\int^{\infty}_{-\infty}\!\!\!dx\int^{\infty}_{-\infty}\!\!\!dy~\epsilon(x-y)
j^{R,-}_t(x)\left(\frac{1-C}{\sqrt{C}}j^{R,3}_t-2ij^{R,3}_x\right)(y) \nonumber \\
&&+2i\frac{1+C}{\sqrt{C}}\int^{\infty}_{-\infty}\!\!\!dx~j^{R,-}_x(x) 
-\frac{1}{6}(Q^{R,-}_{(1)})^2\widetilde{Q}^{R,+}_{(1)}-\frac{1-C^2}{C}Q^{R,-}_{(1)}\,, \nonumber \\
\widetilde{Q}^{R,+}_{(3)} &=& \frac{1}{2}\int^{\infty}_{-\infty}\!\!\!dx\int^{\infty}_{-\infty}\!\!\!dy
\int^{\infty}_{-\infty}\!\!\!dz~\epsilon(x-y)\epsilon(x-z)j^{R,-}_t(x)\widetilde{j}^{R,+}_t(y)\widetilde{j}^{R,+}_t(z) \nonumber \\
&&+\int^{\infty}_{-\infty}\!\!\!dx\int^{\infty}_{-\infty}\!\!\!dy~\epsilon(x-y)
\widetilde{j}^{R,+}_t(x)\left(\frac{1-C}{\sqrt{C}}j^{R,3}_t - 2i j^{R,3}_x \right)(y) \nonumber \\
&&+2i\frac{1+C}{\sqrt{C}}\int^{\infty}_{-\infty}\!\!\!dx~\widetilde{j}^{R,+}_x(x) 
-\frac{1}{6}(\widetilde{Q}^{R,+}_{(1)})^2 Q^{R,-}_{(1)}-\frac{1-C^2}{C}\widetilde{Q}^{R,+}_{(1)}\,, \nonumber \\ 
&& \qquad \vdots \label{list+}
\end{eqnarray}
The subscript $(n)$ of $Q_{(n)}$, which we call level, 
denotes the order of $z$ and also measures the non-locality of the charges 
simultaneously. We have introduced the signature function $\epsilon(x-y) \equiv \theta(x-y) - \theta(y-x)$\,,  
where $\theta(x-y)$ is a step function. 

\medskip 

Note that all of the charges in (\ref{list+}) are written in terms of non-local currents,\footnote{
The appearance of non-local currents is suggested also from the T-duality argument \cite{ORU}.} 
\begin{eqnarray}
j^{R,3}_{\mu}(x) &\equiv&(1+C)J^3_{\mu}(x) \qquad \mbox{(local)}\,, 
\nonumber \\
j^{R,\pm}_{\mu}(x) &\equiv&{\rm e}^{\gamma\chi}\left[J^{\pm}_{\mu}\pm i\sqrt{C}\epsilon_{\mu\nu}J^{\pm,\nu}\right](x)\,, 
\label{NLcurrent} \\
\widetilde{j}^{R,\pm}_{\mu}(x) &\equiv&{\rm e}^{-\gamma\chi}\left[J^{\pm}_{\mu}\mp i\sqrt{C}\epsilon_{\mu\nu}J^{\pm,\nu}\right](x)\,.  
\nonumber \\ 
\chi(x) &\equiv& \frac{1}{2}\int^{\infty}_{-\infty}\!\!\!dy~\epsilon(x-y)j^{R,3}_t(y) \qquad \mbox{(non-local)}\,. \nonumber 
\end{eqnarray}
This point is highly non-trivial because the Lax pair is not written in terms of the non-local currents in (\ref{NLcurrent}) 
but the left-invariant current $J=g^{-1}dg$\,. Then a direct computation shows that 
all of the currents in (\ref{NLcurrent}) are conserved under the equations of motion in (\ref{eom}) 
and the corresponding conserved charges can be constructed. 
In fact, in the previous work \cite{KYhybrid}, we have already found out the first three currents  
$j_{\mu}^{R,3}$ and $j_{\mu}^{R,\pm}$ and have shown that the corresponding charges generate 
a quantum group algebra $U_q(sl(2))$\,. This is a non-local realization of 
the broken $SU(2)_{\rm R}$ generators according to the squashing of the target space geometry. 

\medskip 

The remaining question is what is the role of new ingredients $\widetilde{j}_{\mu}^{R,\pm}$\,. 
As we will discuss later, the corresponding charges enhance $U_q(sl(2))$ to a classical analogue of quantum affine algebra 
$U_q(\widehat{sl(2)})$\,. That is, $\widetilde{j}_{\mu}^{R,\pm}$ are related to its affine generators. 

\medskip 

Finally we should notice that the conserved charges, 
\[
Q_{(1)}^{R,+} \equiv \int^{\infty}_{-\infty}\!\!\!dx\,j_t^{R,+}\,, \qquad 
\widetilde{Q}_{(1)}^{R,-} \equiv \int^{\infty}_{-\infty}\!\!\!dx\,\widetilde{j}_t^{R,-}
\]
are missed 
in the list (\ref{list+}). This observation suggests that the expansion i) is not enough to consider 
the underlying symmetry of the system, although a single point expansion of the monodromy matrix is enough in the case of Yangians. 
Indeed, the remaining charges appear in the expansion ii)\,, as shown in the next subsection.

\subsection{Expansion ii)}

Next we will consider the expansion ii)\,. The spatial component of the Lax pair is expanded in terms of $z'\equiv 1/z$ as 
\begin{eqnarray}
L^R_x(x;\lambda)&=&-i\sqrt{C}T_3J^3_t(x) \nonumber \\
&& + z'\left[T_+\left( \frac{2C}{1+C}J^+_x(x)  -\frac{2i\sqrt{C}}{1+C}J^+_t(x)  \right)
\right. \nonumber \\ && \left.
\qquad + T_-\left( \frac{2C}{1+C}J^-_x(x)  -\frac{2i\sqrt{C}}{1+C}J^-_t(x) \right)\right]\nonumber \\
&& + z'^2T_3\left(-\frac{2i\sqrt{C}(1-C)}{1+C}J^3_t(x)+\frac{4C}{1+C}J^3_x(x)\right) \nonumber \\
&& + z'^3\left[T_+\left(-\frac{2i\sqrt{C}(1-3C)}{(1+C)^2}J^+_t(x)-\frac{2i\sqrt{C}(3-C)}{(1+C)^2}J^+_x(x)\right)\right. \nonumber \\
&& \qquad \left. + T_-\left(-\frac{2i\sqrt{C}(1-3C)}{(1+C)^2}J^-_t(x)-\frac{2i\sqrt{C}(3-C)}{(1+C)^2}J^-_x(x)\right)\right] 
+ {\mathcal O}(z'^4)\,. \nonumber 
\end{eqnarray}
Then the expanded monodromy matrix is 
\begin{eqnarray}
U^R(\lambda)&=&{\rm P}\exp\left[-\int^{\infty}_{-\infty}\!\!\!dxL^R_x(x;\lambda)\right] \nonumber \\
&=&{\rm e}^{u_0}\left[1+z'u_1 + z'^2\left(u_2 + \frac{1}{2}(u_1)^2\right) \right. \nonumber \\ 
&& \left. +z'^3\left(u_3+\frac{1}{2}(u_2 u_1 + u_1 u_2) + \frac{1}{6}(u_1)^3\right)+{\mathcal O}(z'^4)\right]\,, \nonumber 
\end{eqnarray}
where $u_i~(i=0,1,2,3,\ldots)$ are 
\begin{eqnarray}
&& u_0 = i\gamma T_3Q^{R,3}_{(0)}\,, \qquad 
u_1 = 2i\gamma\left(T_+{\rm e}^{\gamma Q^{R,3}_{(0)}/2}Q^{R,+}_{(1)} + T_-{\rm e}^{-\gamma Q^{R,3}_{(0)}/2}
\widetilde{Q}^{R,-}_{(1)}\right)\,, \nonumber \\
&& u_2 = -2i\gamma^2T_3 Q^{R,3}_{(2)}\,, \qquad 
u_3 = -2i\gamma^3\left(T_+{\rm e}^{\gamma Q^{R,3}_{(0)}/2}Q^{R,+}_{(3)} + T_-{\rm e}^{-\gamma Q^{R,3}_{(0)}/2}
\widetilde{Q}^{R,-}_{(3)}\right)\,, \quad \cdots\,.  \nonumber 
\end{eqnarray}
The conserved charges obtained up to the fourth order of $z'$ are listed below,
\begin{eqnarray}
Q^{R,3}_{(0)} &=& \int^{\infty}_{-\infty}\!\!\!dx~j^{R,3}_t(x) = - \bar{Q}_{(0)}^{R,3}\,, \nonumber \\ 
Q^{R,+}_{(1)} &=& \int^{\infty}_{-\infty}\!\!\!dx~j^{R,+}_t(x)\,, \qquad 
\widetilde{Q}^{R,-}_{(1)} = \int^{\infty}_{-\infty}\!\!\!dx~\widetilde{j}^{R,-}_t(x)\,, \nonumber \\
Q^{R,3}_{(2)} &=& \int^{\infty}_{-\infty}\!\!\!dx\int^{\infty}_{-\infty}\!\!\!dy~\epsilon(x-y)
j^{R,+}_t(x)\widetilde{j}^{R,-}_t(y) \nonumber \\
&& -2i \int^{\infty}_{-\infty}\!\!\!dx\,j^{R,3}_x(x)-\frac{1-C}{\sqrt{C}}Q^{R,3}_{(0)}\,, \nonumber \\
Q^{R,+}_{(3)}&=&\frac{1}{2}\int^{\infty}_{-\infty}\!\!\!dx\int^{\infty}_{-\infty}\!\!\!dy
\int^{\infty}_{-\infty}\!\!\!dz~\epsilon(x-y)\epsilon(x-z)
\widetilde{j}^{R,-}_t(x)j^{R,+}_t(y)j^{R,+}_t(z) \nonumber \\
&&-\int^{\infty}_{-\infty}\!\!\!dx\int^{\infty}_{-\infty}\!\!\!dy~\epsilon(x-y)
j^{R,+}_t(x)\left(\frac{1-C}{\sqrt{C}}j^{R,3}_t+2ij^{R,3}_x\right)(y) \nonumber \\
&&-2i\frac{1+C}{\sqrt{C}}\int^{\infty}_{-\infty}\!\!\!dx~j^{R,+}_x(x) 
-\frac{1}{6}(Q^{R,+}_{(1)})^2 \widetilde{Q}^{R,-}_{(1)}-\frac{1-C^2}{C}Q^{R,+}_{(1)}\,, \nonumber \\
\widetilde{Q}^{R,-}_{(3)}&=&\frac{1}{2}\int^{\infty}_{-\infty}\!\!\!dx\int^{\infty}_{-\infty}\!\!\!dy
\int^{\infty}_{-\infty}\!\!\!dz~\epsilon(x-y)\epsilon(x-z)j^{R,+}_t(x)\widetilde{j}^{R,-}_t(y)\widetilde{j}^{R,-}_t(z) \nonumber \\
&&+\int^{\infty}_{-\infty}\!\!\!dx\int^{\infty}_{-\infty}\!\!\!dy~\epsilon(x-y)
\widetilde{j}^{R,-}_t(x)\left(\frac{1-C}{\sqrt{C}}j^{R,3}_t+2ij^{R,3}_x\right)(y) \nonumber \\
&&-2i\frac{1+C}{\sqrt{C}}\int^{\infty}_{-\infty}\!\!\!dx~\widetilde{j}^{R,-}_x(x) 
-\frac{1}{6}(\widetilde{Q}^{R,-}_{(1)})^2Q^{R,+}_{(1)}-\frac{1-C^2}{C} \widetilde{Q}^{R,-}_{(1)}\,, \nonumber \\ 
&& \qquad \vdots \label{list-} 
\end{eqnarray}
Note that all of the charges are again written in terms of the non-local currents in (\ref{NLcurrent})\,. 
As mentioned in the previous subsection, $Q_{(1)}^{R,+}$ and $\widetilde{Q}_{(1)}^{R,-}$ are surely contained as the first two 
in the list (\ref{list-}). The next task is to clarify the algebraic structure that the conserved charges form.

\subsection{Poisson brackets of non-local charges}

\begin{figure}[htbp]
 \begin{center}
  \includegraphics[scale=.4]{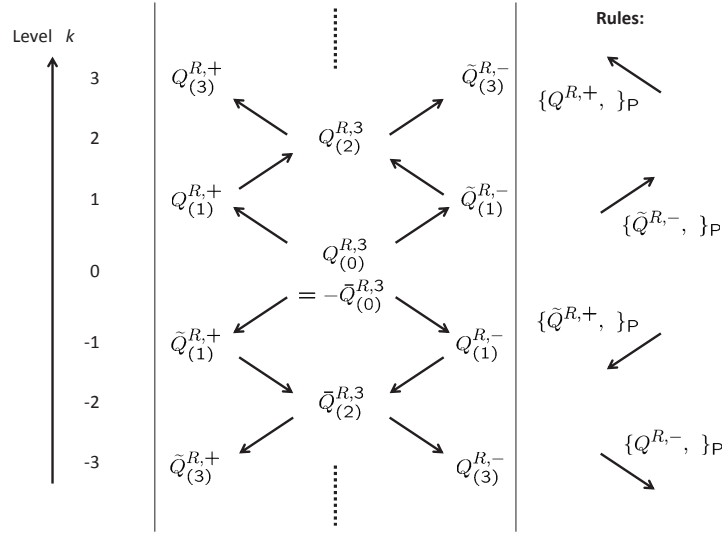}
 \end{center}
\vspace*{-1cm} 
 \caption{\footnotesize The tower of conserved charges. The absolute value of the level 
in the vertical axis measures the non-locality of the charges. The horizontal axis denotes eigenvalues of $Q^{R,3}_{(0)}$\,. 
Higher-level charges can be constructed basically by taking the Poisson bracket according to the rules depicted 
in the figure, up to lower-level conserved charges. 
\label{tower:fig}}
\end{figure}

It is a turn to compute the Poisson brackets of the non-local conserved charges. 
The starting point is the Poisson brackets that the left-invariant one-form $J = g^{-1} dg$ satisfy,    
\begin{eqnarray}
\left\{J^{\pm}_t(x),J^{\mp}_t(y)\right\}_{\rm P}&=&\pm i(1+C)J^3_t(x)\delta(x-y)\,, \nonumber \\ 
\left\{J^{\pm}_t(x),J^3_t(y)\right\}_{\rm P}&=&\mp\frac{1}{1+C}iJ^{\pm}_t(x)\delta(x-y)\,, \nonumber \\ 
\left\{J^{\pm}_t(x),J^{\mp}_x(y)\right\}_{\rm P}&=&\pm iJ^3_x(x)\delta(x-y)+\partial_x\delta(x-y)\,, \nonumber \\ 
\left\{J^{\pm}_t(x),J^3_x(y)\right\}_{\rm P}&=&\mp iJ^{\pm}_x(x)\delta(x-y)\,, \label{c-current} \\ 
\left\{J^3_t(x),J^{\pm}_x(y)\right\}_{\rm P}&=&\pm i\frac{1}{1+C}J^{\pm}_x(x)\delta(x-y)\,, \nonumber \\ 
\left\{J^3_t(x),J^3_x(y)\right\}_{\rm P}&=&\frac{1}{1+C}\partial_x\delta(x-y)\,. \nonumber 
\end{eqnarray}
The relations in (\ref{c-current}) lead to the Poisson brackets of the non-local currents in (\ref{NLcurrent}), 
\begin{eqnarray}
\left\{j^{R,\pm}_t(x),j^{R,\mp}_t(y)\right\}_{\rm P}&=&\pm i{\rm e}^{2\gamma\chi}j^{R,3}_t(x)\delta(x-y)\,, 
\nonumber \\
\left\{j^{R,\pm}_t(x),j^{R,\pm}_t(y)\right\}_{\rm P}&=&\pm i\frac{\sqrt{C}}{1+C}\epsilon(x-y)j^{R,\pm}_t(x)j^{R,\pm}_t(y)\,, 
\nonumber \\
\left\{j^{R,\pm}_t(x),j^{R,3}_t(y)\right\}_{\rm P}&=&\mp ij^{R,\pm}_t(x)\delta(x-y)\,, \nonumber \\
\left\{\widetilde{j}^{R,\pm}_t(x),\widetilde{j}^{R,\mp}_t(y)\right\}_{\rm P}&=& \pm i{\rm e}^{-2\gamma\chi}j^{R,3}_t(x)\delta(x-y)\,, 
\nonumber \\
\left\{\widetilde{j}^{R,\pm}_t(x),\widetilde{j}^{R,\pm}_t(y)\right\}_{\rm P}&=& \mp i\frac{\sqrt{C}}{1+C}\epsilon(x-y)\widetilde{j}^{R,\pm}_t(x)\widetilde{j}^{R,\pm}_t(y)\,, \nonumber  \\
\left\{\widetilde{j}^{R,\pm}_t(x),j^{R,3}_t(y)\right\}_{\rm P}&=& \mp i\widetilde{j}^{R,\pm}_t(x)\delta(x-y)\,, \nonumber \\
\left\{j^{R,\pm}_t(x),\widetilde{j}^{R,\pm}_t(y)\right\}_{\rm P} &=& 0\,, 
\nonumber \\
\left\{j^{R,\pm}_t(x),\widetilde{j}^{R,\mp}_t(y)\right\}_{\rm P} &=& \pm i\frac{1-C}{1+C}j^{R,3}_t(x)\delta(x-y)-\frac{2\sqrt{C}}{1+C}j^{R,3}_x(x)\delta(x-y) \nonumber  \\
&& \mp i\frac{\sqrt{C}}{1+C}\epsilon(x-y)j^{R\pm}_t(x)\widetilde{j}^{R,\mp}_t(y)\pm 2i\sqrt{C}\partial_x\delta(x-y)\,, \nonumber \\
\left\{j^{R,\pm}_t(x),j^{R,3}_x(y)\right\}_{\rm P}&=&\mp ij^{R,\pm}_x(x)\delta(x-y)\,, \nonumber \\
\left\{\widetilde{j}^{R,\pm}_t(x),j^{R,3}_x(y)\right\}_{\rm P} &=& \mp i\widetilde{j}^{R,\pm}_x(x)\delta(x-y)\,, \nonumber  \\
\left\{j^{R,\pm}_t(x),j^{R,\pm}_x(y)\right\}_{\rm P}&=&\pm i\frac{\sqrt{C}}{1+C}\epsilon(x-y)j^{R,\pm}_t(x)j^{R,\pm}_x(y)\,, 
\nonumber \\
\left\{\widetilde{j}^{R,\pm}_t(x),\widetilde{j}^{R,\pm}_x(y)\right\}_{\rm P} &=& 
\mp i\frac{\sqrt{C}}{1+C}\epsilon(x-y)\widetilde{j}^{R,\pm}_t(x)\widetilde{j}^{R,\pm}_x(y)\,. \nonumber
\end{eqnarray}
Integrating this current algebra leads to the following charge algebra, 
\begin{eqnarray}
\left\{Q^{R,\pm}_{(1)},Q^{R,\mp}_{(1)}\right\}_{\rm P}&=& 
\pm i\frac{{\rm e}^{\gamma Q^{R,3}_{(0)}}-{\rm e}^{-\gamma Q^{R,3}_{(0)}}}{2\gamma}\,,  \nonumber \\
\left\{Q^{R,\pm}_{(1)},Q^{R,\pm}_{(1)}\right\}_{\rm P}&=&0\,, \nonumber  \\
\left\{Q^{R,\pm}_{(1)},Q^{R,3}_{(0)}\right\}_{\rm P}&=&\mp iQ^{R,\pm}_{(1)}\,, \nonumber  \\
\left\{\widetilde{Q}^{R,\pm}_{(1)},\widetilde{Q}^{R,\mp}_{(1)}\right\}_{\rm P} &=& 
\pm i\frac{{\rm e}^{\gamma Q^{R,3}_{(0)}}-{\rm e}^{-\gamma Q^{R,3}_{(0)}}}{2\gamma}\,,  \nonumber \\
\left\{\widetilde{Q}^{R,\pm}_{(1)},\widetilde{Q}^{R,\pm}_{(1)}\right\}_{\rm P} &=& 0\,, \label{Poisson nonlocal} \\
\left\{\widetilde{Q}^{R,\pm}_{(1)},Q^{R,3}_{(0)}\right\}_{\rm P} &=& \mp i\widetilde{Q}^{R,\pm}_{(1)}\,,  \nonumber \\
\left\{Q^{R,\pm}_{(1)},\widetilde{Q}^{R,\pm}_{(1)}\right\}_{\rm P} &=& 0\,, \nonumber \\
\left\{Q^{R,+}_{(1)},\widetilde{Q}^{R,-}_{(1)}\right\}_{\rm P}&=&-i\gamma Q^{R,3}_{(2)}\,, \nonumber  \\
\left\{Q^{R,-}_{(1)},\widetilde{Q}^{R,+}_{(1)}\right\}_{\rm P}&=&i\gamma \bar{Q}^{R,3}_{(2)}\,, \nonumber  \\
\left\{Q^{R,3}_{(2)},Q^{R,+}_{(1)}\right\}_{\rm P} &=& 
i\gamma \left[Q^{R,+}_{(3)} + \frac{2}{3}\widetilde{Q}^{R,-}_{(1)}(Q^{R,+}_{(1)})^2\right]\,, \nonumber \\
\left\{Q^{R,3}_{(2)},\widetilde{Q}^{R,-}_{(1)}\right\}_{\rm P} &=& 
-i\gamma \left[\widetilde{Q}^{R,-}_{(3)}+\frac{2}{3}Q^{R,+}_{(1)}(\widetilde{Q}^{R,-}_{(1)})^2\right]\,, \nonumber  \\
\left\{\bar{Q}^{R,3}_{(2)},Q^{R,-}_{(1)}\right\}_{\rm P} &=& 
-i\gamma \left[Q^{R,-}_{(3)} + \frac{2}{3}\widetilde{Q}^{R,+}_{(1)}(Q^{R,-}_{(1)})^2\right]\,, \nonumber \\
\left\{\bar{Q}^{R,3}_{(2)},\widetilde{Q}^{R,+}_{(1)}\right\}_{\rm P} &=& 
i\gamma \left[\widetilde{Q}^{R,+}_{(3)}+\frac{2}{3}Q^{R,-}_{(1)}(\widetilde{Q}^{R,+}_{(1)})^2\right]\,, \nonumber  \\
\left\{Q^{R,+}_{(3)},Q^{R,+}_{(1)}\right\}_{\rm P}&=&\frac{i\gamma}{3}Q^{R,3}_{(2)}(Q^{R,+}_{(1)})^2\,, \nonumber \\
\left\{\widetilde{Q}^{R,-}_{(3)},\widetilde{Q}^{R,-}_{(1)}\right\}_{\rm P} &=&\frac{i\gamma}{3}Q^{R,3}_{(2)}(\widetilde{Q}^{R,-}_{(1)})^2\,, \nonumber \\ 
\left\{Q^{R,-}_{(3)},Q^{R,-}_{(1)}\right\}_{\rm P}&=&-\frac{i\gamma}{3}\bar{Q}^{R,3}_{(2)}(Q^{R,-}_{(1)})^2\,, \nonumber \\
\left\{\widetilde{Q}^{R,+}_{(3)},\widetilde{Q}^{R,+}_{(1)}\right\}_{\rm P} &=&-\frac{i\gamma}{3}\bar{Q}^{R,3}_{(2)}(\widetilde{Q}^{R,+}_{(1)})^2\,, \nonumber \\ 
&& \vdots \nonumber
\end{eqnarray}
where we have used the boundary condition (\ref{bc}) when integrating the first and the fourth brackets. 
Note that higher-level charges can be basically generated by taking the Poisson bracket with $Q^{R,\pm}$ and $\widetilde{Q}^{R,\pm}$\,, 
repeatedly, up to lower-level charges. 
These Poisson brackets enable us to argue the tower structure that the conserved charges form, 
as depicted in Fig.\ \ref{tower:fig}. In fact, this tower can be reinterpreted as the Drinfeld second realization 
of quantum affine algebra $U_q(\widehat{sl(2)})$\,, as we will discuss in the next section. 

\subsection{Yangian limit}

Since both the non-local currents $j_\mu^{R,\pm}(x)$ and $\tilde{j}_\mu^{R,\mp}(x)$ in \eqref{NLcurrent} 
reduce to the local current $J^\pm_\mu(x)$ in $C\to0$ limit, 
it is worth showing how the $SU(2)_{\rm R}$ Yangian charges obtained in \cite{KY} 
are reproduced in this limit. 

\medskip 

Interestingly, we have found that the rescaled differences of the corresponding charges 
recover the $(+,-)$-components of the $SU(2)_{\rm R}$ Yangian generators at level 1 recover as 
\begin{eqnarray}
\label{Ypm}
&& \lim_{C\to 0} \frac{1}{2i\sqrt{C}}\left(Q^{R,+}_{(1)} - \widetilde{Q}^{R,+}_{(1)}\right) 
= \int \!\! dx\,J_x^+(x) - \frac{i}{2}\int\!\!\!\!\int\!\!dxdy\,\epsilon(x-y)\,J_t^+(x)J_t^3(y)\,, \nonumber \\
&& \lim_{C\to 0} \frac{1}{2i\sqrt{C}}\left(\widetilde{Q}^{R,-}_{(1)} - Q^{R,-}_{(1)}\right) 
= \int \!\! dx\,J_x^-(x) + \frac{i}{2}\int\!\!\!\!\int\!\!dxdy\,\epsilon(x-y)\,J_t^-(x)J_t^3(y)\,.  
\end{eqnarray} 
The 3-component of the level 1 Yangian generators is reproduced as the $C \to 0$ limit 
of the difference of $Q_{(2)}^{R,3}$ and $\bar{Q}_{(2)}^{R,3}$
\begin{eqnarray}
\lim_{C\to 0} \frac{i}{4}\left(Q^{R,3}_{(2)} - \bar{Q}_{(2)}^{R,3}\right) 
&=&  \int \!\! dx\,J_x^3(x) + \frac{i}{2}\int\!\!\!\!\int\!\!dxdy\,\epsilon(x-y)\,J_t^+(x)J_t^-(y)\,. 
\end{eqnarray}
Higher-level generators of the $SU(2)_{\rm R}$ Yangian are reproduced similarly. 

\medskip 

In general, the level $n$ generators are obtained as the $C\to 0$ limit of the differences 
$Q^{R,+}_{(2n-1)} - \widetilde{Q}_{(2n-1)}^{R,+}$\,, $\widetilde{Q}^{R,-}_{(2n-1)} - Q_{(2n-1)}^{R,-}$
and $Q^{R,3}_{(2n)} - \bar{Q}_{(2n)}^{R,3}$ for $n\geq 1$\,. That is, half of the tower structure in Fig.\,\ref{tower:fig} 
results in the $SU(2)_{\rm R}$ Yangian after taking the $C \to 0$ limit.

\section{The classical origin of quantum affine algebra}

In this section we will make some interpretations of the Poisson bracket algebra from the mathematical point of view. 
The first thing is that the Poisson brackets of the level $0,\pm1$ charges in the previous section can be regarded as Drinfeld's  
first realization of quantum affine algebra \cite{Drinfeld}. Then we argue the role of the higher-level conserved charges 
in the context of the Drinfeld second realization \cite{Drinfeld}.

\subsection{Drinfeld's first realization of quantum affine algebra}

To see the relation to Drinfeld's first realization \cite{Drinfeld}, 
let us concentrate on the conserved charges $Q_{(0)}^{R,3}$\,, $Q_{(1)}^{R,\pm}$ and $\widetilde{Q}_{(1)}^{R,\pm}$\,,  
apart from the higher-level conserved charges $Q_{(n)}~(n\geq 2)$\,. 
The role of the higher-level charges will be the subject in the next subsection. 

\medskip 

It is convenient to rewrite the charges $Q_{(0)}^{R,3}$\,, $Q_{(1)}^{R,\pm}$ and $\widetilde{Q}_{(1)}^{R,\pm}$ 
as follows:\footnote{We follow the notation utilized in \cite{CP}.}
\[
\begin{array}{ll} 
H_1\equiv -2Q^{R,3}_{(0)}\,, & ~~\qquad H_0\equiv 2Q^{R,3}_{(0)}\,, \notag \\
E_1\equiv \left(\frac{\gamma}{\sinh(\gamma/2)}\right)^{1/2}Q^{R,+}_{(1)}\,, & ~~\qquad 
E_0\equiv \left(\frac{\gamma}{\sinh(\gamma/2)}\right)^{1/2}\widetilde{Q}^{R,-}_{(1)}\,, \notag \\
F_1\equiv \left(\frac{\gamma}{\sinh(\gamma/2)}\right)^{1/2}Q^{R,-}_{(1)}\,, & ~~\qquad 
F_0\equiv \left(\frac{\gamma}{\sinh(\gamma/2)}\right)^{1/2}\widetilde{Q}^{R,+}_{(1)}\,. \notag 
\end{array}
\]
The Poisson brackets of them are  
\begin{eqnarray}
&& i\left\{H_i,H_j\right\}_{\rm P}=0 \qquad (i,j=0,1)\,, \nonumber \\
&& i\left\{H_i,E_j\right\}_{\rm P}= A_{ij}E_j\,,\qquad
i\left\{H_i,F_j\right\}_{\rm P}=-A_{ij}F_j\,, \label{defining}  \\ 
&& i\left\{E_i,F_j\right\}_{\rm P} = \delta_{ij}\frac{q^{H_i}-q^{-H_i}}{q-q^{-1}}\,. \nonumber 
\end{eqnarray}
Here the generalized Cartan matrix $A_{ij}$ is given by 
\begin{eqnarray}
A_{ij}= (\alpha_i,\alpha_j) = 
\begin{pmatrix}
2 & -2 \\ 
-2 & 2 
\end{pmatrix}
\quad \mbox{with} \quad \alpha_1 = 
\begin{pmatrix}
1 \\ 
-1 
\end{pmatrix}
\,, ~~ 
\alpha_0 = 
\begin{pmatrix}
-1 \\ 
1 
\end{pmatrix}
\end{eqnarray}
and a $q$-deformation parameter is defined as
\begin{eqnarray}
q\equiv{\rm e}^{\gamma/2}=\exp\left(\frac{\sqrt{C}}{2(1+C)}\right)\,. 
\end{eqnarray}
The brackets in (\ref{defining}) give a classical realization of the defining relations of quantum affine algebra 
in the sense of the first realization by Drinfeld \cite{Drinfeld}. Its affine central charge $k$ is zero because 
\[
 k \equiv H_0 + H_1 = 0\,. 
\] 
This corresponds to the evaluation representation of quantum affine algebra (see also \cite{CP}). 
Note that the $C\to 0$ limit is equivalent to $q \to 1$\,.

\medskip 

The $q$-Serre relations should also be checked. The classical analogue of the $q$-Serre relations 
are deduced by introducing the classical $q$-Poisson bracket,
\begin{eqnarray}
\left\{J^A,J^B\right\}_{q{\rm P}} \equiv\left\{J^A,J^B\right\}_{\rm P}+\frac{i\gamma}{2}\left(\beta_A,\beta_B\right)J^B J^A\,, 
\label{q-Poisson}
\end{eqnarray}
where $\beta_A$ are the associated root vectors. Now $J^A$ and $J^B$ are $c$-number and commutative and 
the ordering in the second term is irrelevant. 
This $q$-Poisson bracket in (\ref{q-Poisson}) is nothing but a classical analogue of $q$-commutator 
and it is realized as a semiclassical limit ($\hbar \to 0$) of the $q$-commutator at quantum level, as we will see later.  

\medskip 

With the $q$-Poisson bracket in (\ref{q-Poisson}), the classical $q$-Serre relations are shown as 
\begin{eqnarray}
\bigl\{E_i, \bigl\{E_i, \bigl\{E_i, E_j \bigr\}_{q{\rm P}} \bigr\}_{q{\rm P}} \bigr\}_{q{\rm P}} = 
\bigl\{F_i, \bigl\{F_i, \bigl\{F_i, F_j \bigr\}_{q{\rm P}} \bigr\}_{q{\rm P}} \bigr\}_{q{\rm P}} = 0 \nonumber 
\quad {\rm for}~~ |i-j|=1\,. 
\label{q-serre}
\end{eqnarray}
For the detail computation, see Appendix \ref{proof}.

\subsection{The relation to the second realization}

Next we shall make an interpretation of the higher-level conserved charges in the context of the Drinfeld 
second realization of quantum affine algebra \cite{Drinfeld}. 

\medskip 

Let us first introduce the following notation, 
\begin{eqnarray}
&& h_0 \equiv -2Q^{R,3}_{(0)}\,, \qquad \quad x_0^+ \equiv \sqrt{2}\,Q^{R,+}_{(1)}\,, \qquad \quad
x_0^- \equiv \sqrt{2}\,Q^{R,-}_{(1)}\,, \nonumber \\
&& x_{-1}^+ \equiv \sqrt{2}\,{\rm e}^{\gamma Q^{R,3}_{(0)}}\, \widetilde{Q}^{R,+}_{(1)}\,, \qquad \quad 
x_1^- \equiv \sqrt{2}\,{\rm e}^{\gamma \bar{Q}^{R,3}_{(0)}}\, \widetilde{Q}^{R,-}_{(1)}\,. \label{second}
\end{eqnarray}
Equivalently, the relations between $\left\{h_k,x_k^\pm\right\}_{k\in{\mathbb Z}}$ and $H_i,E_i,F_i ~(i=0,1)$ are written as 
\begin{align}
H_1 &= h_0\,,& 
E_1 &= \left(\frac{\gamma/2}{\sinh(\gamma/2)}\right)^{1/2}x_0^+\,,&
F_1 &= \left(\frac{\gamma/2}{\sinh(\gamma/2)}\right)^{1/2}x_0^-\,,\nonumber\\
H_0 &= -h_0\,,& 
E_0 &= \left(\frac{\gamma/2}{\sinh(\gamma/2)}\right)^{1/2}{\rm e}^{-\gamma h_0/2}\,x_1^-\,,& 
F_0 &= \left(\frac{\gamma/2}{\sinh(\gamma/2)}\right)^{1/2}{\rm e}^{\gamma h_0/2}\,x_{-1}^+\,.
\label{iso}
\end{align} 
This is the isomorphism from the first to the second realizations \cite{Drinfeld} (see also \cite{CP}).  

\medskip 

\begin{figure}[htbp]
 \begin{center}
  \includegraphics[scale=.3]{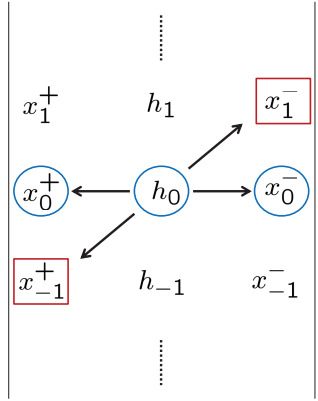}
 \end{center}
\vspace*{-1cm} 
 \caption{\footnotesize The tower structure of the conserved charges in the Drinfeld second realization. 
The three circles denote the $sl(2)$ root diagram and the two boxes are the associated affine generators. 
With the affine generators, higher conserved charges are basically generated according to the composition laws of vectors. 
First of all, $h_1$ is constructed as $h_1 \sim \{x_0^+,x_1^-\}_{\rm P}$\,. Then $x_1^+$ is generated by $h_1$ 
and $x_0^+$ like $x_1^+ \sim \{x_0^+,h_1\}$\,. The next step is to generate $h_2$ with $x_1^+$ and $x_1^-$\,. 
After that $x_2^{\pm}$ are obtained by acting $x_0^{\pm}$ to $h_2$\,, respectively. 
This step can be repeated recursively and the upper half of the tower is generated. The lower half is also generated in the same way. 
\label{Drinfeld2:fig}}
\end{figure}

With the definitions in (\ref{second}) and the Poisson brackets in (\ref{Poisson nonlocal}), 
one can show that the following relations are satisfied, 
\begin{eqnarray}
&&\left\{h_k,h_l\right\}_{\rm P} = 0\,,\quad \left\{h_k,x_l^\pm\right\}_{\rm P} = \mp 2i x_{k+l}^\pm\,,\nonumber\\
&&\left\{x_{k+1}^\pm,x_l^\pm\right\}_{\rm P}\pm i\gamma x_l^\pm x_{k+1}^\pm = \left\{x_k^\pm,x_{l+1}^\pm\right\}_{\rm P}\mp i\gamma x_k^\pm x_{l+1}^\pm\,,\nonumber\\
&&\left\{x_k^+,x_l^-\right\}_{\rm P} = -\frac{i}{\gamma}\left(\psi_{k+l}^+ - \psi_{k+l}^-\right)\,,\nonumber \\
&&\sum_{k\in{\mathbb Z}}\psi_k^\pm z^{-k} = {\rm e}^{\pm\gamma h_0/2}\exp\left(\pm\gamma\sum_{k=1}^\infty h_{\pm k}z^{\mp k}\right)\,.
\end{eqnarray}
This is nothing but a classical analogue of $U_q(\widehat{sl(2)})$ in the sense of the second realization. 
The root diagram of the conserved charges 
is depicted in Fig.\,\ref{Drinfeld2:fig}. 

\medskip 

The explicit expressions of higher charges can be computed from the above relations. 
For example, $h_1$ and $h_{-1}$ are obtained from $\left\{x_0^+,x_1^-\right\}_{\rm P}$ and $\left\{x_{-1}^+,x_0^-\right\}_{\rm P}$\, respectively, 
\begin{eqnarray}
&&\bigl\{x_0^+,x_1^-\bigr\}_{\rm P}=-\frac{i}{\gamma}\left(\psi_1^+-\psi_1^-\right)=-i{\rm e}^{h_0\gamma/2}h_1\,,\nonumber \\
&&\bigl\{x_{-1}^+,x_0^-\bigr\}_{\rm P}=-\frac{i}{\gamma}\left(\psi_{-1}^+-\psi_{-1}^-\right)=-i{\rm e}^{-h_0\gamma/2}h_{-1}\,, \nonumber 
\end{eqnarray}
and hence they can be written in terms of $Q_{(1)}^{R,\pm}$ and $\widetilde{Q}_{(1)}^{R,\pm}$\,, 
\begin{eqnarray}
h_1 &=& 2i{\rm e}^{\gamma Q^{R,3}_{(0)}}\left\{Q^{R,+}_{(1)},{\rm e}^{\gamma \bar{Q}^{R,3}_{(0)}}
\widetilde{Q}^{R,-}_{(1)}\right\} \nonumber \\
&=& 2i\left\{Q^{R,+}_{(1)},\widetilde{Q}^{R,-}_{(1)}\right\}_{\rm P} - 2\gamma Q^{R,+}_{(1)}\widetilde{Q}^{R,-}_{(1)} 
= -2i \left\{\widetilde{Q}^{R,-}_{(1)},Q^{R,+}_{(1)}\right\}_{q\rm P}\,,\nonumber \\
h_{-1} &=& 2i{\rm e}^{\gamma \bar{Q}^{R,3}_{(0)}}\left\{{\rm e}^{\gamma Q^{R,3}_{(0)}}
\widetilde{Q}^{R,+}_{(1)},Q^{R,-}_{(1)}\right\} \nonumber \\
&=& 2i\left\{\widetilde{Q}^{R,+}_{(1)},Q^{R,-}_{(1)}\right\}_{\rm P} + 2\gamma \widetilde{Q}^{R,+}_{(1)}Q^{R,-}_{(1)} 
= 2i \left\{\widetilde{Q}^{R,+}_{(1)},Q^{R,-}_{(1)}\right\}_{q\rm P}\,. \nonumber 
\end{eqnarray}
Then $x_k^\pm~(k=1,2,3,\ldots)$ are constructed as a sequence obtained by acting $h_{\pm 1}$ on $x_0^{\pm}$ repeatedly,  
\begin{eqnarray}
&&\bigl\{h_1,x_k^\pm\bigr\}_{\rm P} = \mp 2ix_{k+1}^\pm \nonumber \\ 
&\Longrightarrow& \qquad 
x_k^\pm = \left(\pm\frac{i}{2}\right)^k\bigl\{h_1,\bigl\{h_1,\bigl\{\cdots,
\bigl\{h_1,x_0^\pm\bigr\}_{\rm P}\bigr\}_{\rm P}\bigr\}_{\rm P}\bigr\}_{\rm P}\,, \nonumber \\
&&\bigl\{h_{-1},x_{-k}^\pm\bigr\}_{\rm P} = \mp 2ix_{-k-1}^\pm \nonumber \\
&\Longrightarrow& \qquad 
x_{-k}^\pm = \left(\pm\frac{i}{2}\right)^k\bigl\{h_{-1},\bigl\{h_{-1},
\bigl\{\cdots,\bigl\{h_{-1},x_0^\pm\bigr\}_{\rm P}\bigr\}_{\rm P}\bigr\}_{\rm P}\bigr\}_{\rm P}\,. \nonumber 
\end{eqnarray}
Since $x_0^{\pm}$\,, $h_1$ and $h_{-1}$ are written in terms of $Q^{R,3}_{(0)}$\,, 
$Q^{R, \pm}_{(1)}$ and $\widetilde{Q}^{R, \pm}_{(1)}$\,, $x_k^{\pm}$ are also written in the same way. 
The Poisson brackets $\left\{A,B\right\}_{\rm P}$ above may be replaced by $q$-Poisson brackets $\left\{A,B\right\}_{q\rm P}$\,,  
because the inner product of the root vectors associated with 
$\bigl\{Q^{R,\pm}_{(1)},\widetilde{Q}^{R,\mp}_{(1)}\bigr\}_{q\rm P}$ vanishes    
and there is no correction term in \eqref{q-Poisson}.
In the end, all of $h_k$ are obtained from $x_k^\pm$ with the relations 
\begin{eqnarray}
&&\left\{x_{k-1}^+,x_1^-\right\}_{\rm P} = \frac{\psi_k^+-\psi_k^-}{i\gamma} = -i{\rm e}^{\gamma h_0/2}h_k + \cdots\,, \nonumber\\
&&\left\{x_{-1}^+,x_{-k+1}^-\right\}_{\rm P} = \frac{\psi_{-k}^+-\psi_{-k}^-}{i\gamma} = -i{\rm e}^{\gamma h_0/2}h_{-k} + \cdots\,. 
\nonumber 
\end{eqnarray}
Here the part ``$\dots$'' contains only products of the lower-level conserved charges. 
The above argument proves the surjectivity of the map \eqref{iso}. 

\medskip 

Let us here comment on the relation between the second realization of quantum affine algebra 
and the higher-level conserved charges obtained by expanding the monodromy matrix $U^R(\lambda)$ in (\ref{monodromy}). 
By construction, $x^{\pm}_{\pm k}$ and $h_{\pm k}$ are written as a sequence of the Poisson brackets 
among $Q^{R,\pm}_{(1)}$ and $\widetilde{Q}^{R,\pm}_{(1)}$\,. 
Hence it is easy to notice that $x^{\pm}_{\pm k}$ and $h_{\pm k}$ are closely related to the higher-level conserved charges 
obtained by expanding the monodromy matrix. For example, $h_1$ and $h_{-1}$ correspond to $Q_{(2)}^{R,3}$ and $\bar{Q}_{(2)}^{R,3}$\,, 
respectively, up to the lower-level conserved charges. Similarly, one can figure out the correspondence between 
the charges in Fig.\ \ref{tower:fig} and in Fig.\ \ref{Drinfeld2:fig}, up to lower-level conserved charges. 

\medskip 

Note that there is an ambiguity in the expression of the monodromy matrix in (\ref{monodromy}) according to an ambiguity 
of the Lax pair due to gauge transformations. It may be possible to figure out the exact 
correspondence without deviation by lower-level conserved charges. However, it has not been done yet so far.    
As a peculiarity of $U_q(\widehat{sl(2)})$\,, the width of the root diagram shown in Fig.\,\ref{Drinfeld2:fig} 
is not so wide that such an exact correspondence may be found out. It would be an interesting direction in the future study. 
It is also nice to elucidate the relation to the RTT formalism, following \cite{DF}.

\section{Comment on semiclassical limit}

Although we have focused upon classical realizations of quantum affine algebra so far,  
the next subject is to consider a semiclassical limit of quantum affine algebra realized at quantum mechanical level. 
In principle, one can perform the canonical quantization 
by replacing the classical Poisson bracket with the usual commutator like  
\[
i\{~~,~~\}_{\rm P} ~~\rightarrow~~ \frac{1}{\hbar}[~~,~~]\,. 
\]
Then a quantum affine algebra seems to be realized at quantum level but it is not the case. 
The conservation laws of non-local charges should be checked carefully, because their definition contains 
the product of currents and hence some renormalizations are necessary to define the charges at quantum level definitely. 
Namely, the conservation laws might be broken due to the renormalization after all. In the case of $O(N)$ non-linear sigma models  
in two dimensions, the quantum conservation laws are carefully confirmed \cite{Luscher1} 
(For generic coset sigma models, see \cite{GW,AFG}). 

\medskip 
 
Eventually, the quantum conservation laws should be shown for definite argument by following \cite{GW,AFG} in the present case. 
Then it is possible to discuss the quantum affine algebra along the scenario as discussed in \cite{Bernard}.  
We do not, however, try to argue the conservation laws in detail here and leave it as a future problem. 
Instead, simply supposing that well-defined quantum charges $\widehat{Q}$'s exist, 
we discuss a semiclassical limit of quantum affine algebra realized at quantum level. 

\medskip 

Note that, for quantum integrability of squashed sigma model, we have another confirmation, which is that 
the Bethe ansatz has already been constructed by Wiegmann \cite{quantum1} (For related works see \cite{quantum2,quantum3}) 
and the exact solutions have been found out. As a result, the quantum integrability has been confirmed indirectly by another argument. 

\medskip 

For simplicity, we consider the first realization of quantum affine algebra here. 
Then the quantum charges satisfy the defining relations of $U_q(\widehat{sl(2)})$\,, which are the standard form 
in mathematical literatures, like 
\begin{eqnarray}
&& \bigl[\widehat{H}_i,\widehat{H}_j\bigr] = 0 \qquad (i,j=0,1)\,, \nonumber \\
&& \bigl[\widehat{H}_i,\widehat{E}_j\bigr] = A_{ij}\widehat{E}_j\,, \qquad 
\bigl[\widehat{H}_i,\widehat{F}_j\bigr] = -A_{ij}\widehat{F}_j\,, \label{qaffine} \\
&& \bigl[\widehat{E}_i,\widehat{F}_j\bigr] = \delta_{ij}\frac{\widehat{q}^{H_i}-\widehat{q}^{-H_i}}{\widehat{q}-\widehat{q}^{-1}}\,.  \nonumber
\end{eqnarray}
The $q$-Serre relations are 
\begin{eqnarray}
\bigl[\widehat{E}_i,\bigl[\widehat{E}_i,\bigl[\widehat{E}_i,\widehat{E}_j\bigr]_{\widehat{q}}\bigr]_{\widehat{q}}\bigr]_{\widehat{q}} = 
\bigl[\widehat{F}_i,\bigl[\widehat{F}_i,\bigl[\widehat{F}_i,\widehat{F}_j\bigr]_{\widehat{q}}\bigr]_{\widehat{q}}\bigr]_{\widehat{q}} = 0
\qquad {\rm for} \quad |i-j|=1\
\label{qserre}
\end{eqnarray}
and the $q$-commutator is defined as
\begin{eqnarray}
[\widehat{J}^A,\widehat{J}^B]_{\widehat{q}} \equiv \widehat{J}^A \widehat{J}^B 
- \widehat{q}^{(\beta_A,\beta_B)}\widehat{J}^B \widehat{J}^A\,. 
\label{q-commutator}
\end{eqnarray}
Here a deformation parameter $\widehat{q}$ at quantum level is related to the classical one $q$ as
\begin{eqnarray}
\widehat{q}\equiv q^\hbar={\rm e}^{\hbar\gamma/2}\,.
\end{eqnarray}
Note that $\widehat{q}$ depends on the Planck constant $\hbar$\,. This is a difference of importance between at classical and 
quantum levels.

\medskip 

Let us now consider a semiclassical limit $\hbar \to 0$\,. 
The quantum charges are first rescaled as 
\begin{eqnarray} 
\widehat{E}_i\rightarrow\hbar\left(\frac{\gamma}{\sinh(\gamma/2)}\right)^{1/2}\widehat{E}_i\,, \quad 
\widehat{F}_i\rightarrow\hbar\left(\frac{\gamma}{\sinh(\gamma/2)}\right)^{1/2}\widehat{F}_i\,, \quad  
\widehat{H}_i\rightarrow\frac{\hbar}{2}\widehat{H}_i\,, \nonumber
\end{eqnarray}
and then the commutators should be replaced by the Poisson brackets, 
\begin{eqnarray}
\left[\quad,\quad\right] ~~\rightarrow~~ i\hbar\left\{\quad,\quad\right\}_{\rm P}\,. \nonumber 
\end{eqnarray}
Noting that $\widehat{q}$ is expanded with respect to $\hbar$ as 
\begin{eqnarray}
\widehat{q}=1+\frac{\hbar\gamma}{2}+\mathcal{O}(\hbar^2)\,, \nonumber 
\end{eqnarray}
the semiclassical limit $\hbar \to 0$ is taken. 

\medskip 

As a result, the classical defining relations in (\ref{defining}) are reproduced 
as a semiclassical limit of the quantum ones in (\ref{qaffine})\,, 
as a matter of course. 
In addition, the classical $q$-Poisson bracket in (\ref{q-Poisson}) is reproduced as a semiclassical limit of 
the standard $q$-commutator (\ref{q-commutator}): 
\begin{eqnarray}
[\widehat{J}^A,\widehat{J}^B]_{\widehat{q}} &=& \widehat{J}^A \widehat{J}^B - \widehat{q}^{(\beta_A,\beta_B)}\widehat{J}^B \widehat{J}^A \nonumber \\ 
&=& [\widehat{J}^A,\widehat{J}^B] -({\rm e}^{\hbar\gamma(\beta_A,\beta_B)/2}-1)\widehat{J}^B \widehat{J}^A \nonumber \\ 
&\rightarrow& i\hbar\{J^A,J^B\}_{\rm P} -\frac{\hbar\gamma}{2} (\beta_A,\beta_B) J^B J^A \nonumber \\ 
&=& i\hbar\{J^A,J^B\}_{q{\rm P}} \nonumber 
\end{eqnarray}

\section{Conclusion and Discussion}

We have argued a quantum affine algebra realized in two-dimensional non-linear sigma models  
with target space three-dimensional squashed spheres. We have explicitly constructed its affine generators 
and computed the Poisson brackets. The defining relations of quantum affine algebra in the sense 
of the Drinfeld first realization are satisfied at classical level. 
The relation to the second realization is also discussed including higher conserved charges. 
The result here is consistently interpreted as a semiclassical limit of quantum affine algebra realized at quantum level. 

\medskip 

There are some potentially interesting directions in the future study. 
The first is to figure out an affine extension of $q$-deformed Poincare symmetry in the null-warped case \cite{KY-Sch} 
by following the argument discussed here. 
It is also nice to consider an extension of the null-warped geometry to the higher-dimensional case,  
though the coset structure is not reductive any more in contrast to the three-dimensional case \cite{SYY}. 
A relative direction is to consider the hybrid deformation consisting of the standard $q$-deformed $SL(2)$ 
and the $q$-deformed Poincare \cite{BHP} (For its application to three-dimensional gravities see \cite{BHM}). 

\medskip 

The second is to look for some applications in the context of AdS/condensed matter physics (CMP), 
where the warped AdS geometries appear as the gravity dual to the system in the presence of magnetic field \cite{Kraus}. 
The anisotropy of the system is reflected as the squashing of spacetime geometry in the gravity side. 
Finally, it is interesting to try to consider quantum affine algebra in the context of Kerr/CFT correspondence 
\cite{Kerr/CFT} and the recently proposed scenario, warped AdS$_3$/dipole CFT$_2$ \cite{Guica,SS}. 

\medskip 

It is also a nice direction to consider the string-theory embedding by following the works \cite{Detournay1,Detournay2} 
and consider the role of quantum affine algebra presented here in the string-theory context. In this direction,  
first of all, we should be careful for the conformal invariance. The squashed sigma model is not conformal 
and hence we have to add the Wess-Zumino (WZ) term. We have already shown that 
the $SU(2)_{\rm L}$ Yangian algebra is still preserved even after adding the WZ term \cite{KOY}. However,  
the quantum affine algebra in the presence of the WZ term has not been investigated yet. 
It is the next issue and we hope that we could report on it in the near future.

\subsection*{Acknowledgments}

We would like to thank H.~Kawai, S.~Moriyama and T.~Okada for illuminating discussions. This work was initiated in the workshop, 
under the program ``Synthesis of integrabilities arising from gauge-string duality,''  held 
at Higher School of Economics and Steklov Mathematical Institute in Moscow, Russia. We greatly appreciate 
the organizers' hospitality, including H.~Itoyama and A.~Morozov. 
The work of IK was supported by the Japan Society for the Promotion of Science (JSPS). 
The work of KY was supported by the scientific grants from the Ministry of Education, Culture, Sports, Science 
and Technology (MEXT) of Japan (No.\,22740160). 
This work was also supported in part by the Grant-in-Aid 
for the Global COE Program ``The Next Generation of Physics, Spun 
from Universality and Emergence'' from MEXT, Japan. 
One of the authors TM is also grateful to A.~Molev for valuable comments on this work. 
Part of his work was done during ``2nd Asia-Pacific Summer School in Mathematical Physics, 
22nd Canberra International Physics Summer School CFT, AdS/CFT and Integrability"
held at the Australian National University in Canberra, Australia. 
He would like to thank the organizers and the lecturers C.~Ahn, V.~Bazhanov and R.~Nepomechie for 
kindly answering his basic questions related to the subject of this paper.

\appendix 

\section*{Appendix}

\section{Proof of $q$-Serre relations at classical level \label{proof}}

We show here that $E_i$ and $F_i$ satisfy the classical analogue of $q$-Serre relations in (\ref{q-serre}). 
Note that the $q$-Serre relations are rewritten with $Q^{R,\pm}_{(1)}$ and $\widetilde{Q}^{R,\pm}_{(1)}$ as 
\begin{eqnarray}
&&\bigl\{Q^{R,\pm}_{(1)}, \bigl\{Q^{R,\pm}_{(1)}, \bigl\{Q^{R,\pm}_{(1)},\widetilde{Q}^{R,\mp}_{(1)}\bigr\}_{q{\rm P}}\bigr\}_{q{\rm P}}\bigr\}_{q{\rm P}} = 0\,, 
\label{q-serre1} \\
&&\bigl\{\widetilde{Q}^{R,\mp}_{(1)}, \bigl\{\widetilde{Q}^{R,\mp}_{(1)}, \bigl\{\widetilde{Q}^{R,\mp}_{(1)},Q^{R,\pm}_{(1)}\bigr\}_{q{\rm P}}\bigr\}_{q{\rm P}}\bigr\}_{q{\rm P}} = 0\,.
\label{q-serre2}
\end{eqnarray}
The first bracket is evaluated as 
\begin{eqnarray}
\bigl\{Q^{R,\pm}_{(1)},\widetilde{Q}^{R,\mp}_{(1)}\bigr\}_{q{\rm P}} 
&=& \bigl\{Q^{R,\pm}_{(1)},\widetilde{Q}^{R,\mp}_{(1)}\bigr\}_{\rm P} 
+ \frac{i\gamma}{2} \left(\alpha_0,\alpha_1\right)\widetilde{Q}^{R,\mp}_{(1)}Q^{R,\pm}_{(1)} \nonumber \\
&=& \bigl\{Q^{R,\pm}_{(1)},\widetilde{Q}^{R,\mp}_{(1)}\bigr\}_{\rm P} 
- i\gamma \widetilde{Q}^{R,\mp}_{(1)}Q^{R,\pm}_{(1)}\,. \nonumber 
\end{eqnarray}
Then one more bracket leads to 
\begin{eqnarray}
&& \bigl\{Q^{R,\pm}_{(1)}, \bigl\{Q^{R,\pm}_{(1)},\widetilde{Q}^{R,\mp}_{(1)}\bigr\}_{q{\rm P}}\bigr\}_{q{\rm P}} \nonumber\\ 
&=& \bigl\{Q^{R,\pm}_{(1)}, \bigl\{Q^{R,\pm}_{(1)},\widetilde{Q}^{R,\mp}_{(1)}\bigr\}_{q{\rm P}}\bigr\}_{\rm P} 
+ \frac{i\gamma}{2}\left(\alpha_0,\alpha_0+\alpha_1\right)\bigl\{Q^{R,\pm}_{(1)},\widetilde{Q}^{R,\mp}_{(1)}\bigr\}_{q{\rm P}}Q^{R,\pm}_{(1)} \nonumber \\
&=& \bigl\{Q^{R,\pm}_{(1)}, \bigl\{Q^{R,\pm}_{(1)},\widetilde{Q}^{R,\mp}_{(1)} \bigr\}_{q{\rm P}} \bigr\}_{\rm P}\,. \nonumber 
\end{eqnarray}
With one more bracket, we obtain that 
\begin{eqnarray}
&& \bigl\{Q^{R,\pm}_{(1)}, \bigl\{Q^{R,\pm}_{(1)}, \bigl\{Q^{R,\pm}_{(1)}, 
\widetilde{Q}^{R,\mp}_{(1)} \bigr\}_{q{\rm P}} \bigr\}_{q{\rm P}} \bigr\}_{q{\rm P}} \nonumber \\
&=& \bigl\{Q^{R,\pm}_{(1)}, \bigl\{Q^{R,\pm}_{(1)}, \bigl\{Q^{R,\pm}_{(1)}, 
\widetilde{Q}^{R,\mp}_{(1)} \bigr\}_{q{\rm P}} \bigr\}_{q{\rm P}} \bigr\}_{\rm P} \nonumber \\ && \quad 
+\frac{i\gamma}{2}\left(\alpha_0,2\alpha_0+\alpha_1\right)\bigl\{Q^{R,\pm}_{(1)},\bigl\{Q^{R,\pm}_{(1)},\widetilde{Q}^{R,\mp}_{(1)}
\bigr\}_{q{\rm P}}\bigr\}_{q{\rm P}}Q^{R,\pm}_{(1)} \nonumber \\
&=& \bigl\{Q^{R,\pm}_{(1)},\bigl\{Q^{R,\pm}_{(1)},\bigl\{Q^{R,\pm}_{(1)},\widetilde{Q}^{R,\mp}_{(1)} \bigr\}_{q{\rm P}} \bigr\}_{q{\rm P}} \bigr\}_{\rm P} 
+ i\gamma \bigl\{Q^{R,\pm}_{(1)}, \bigl\{Q^{R,\pm}_{(1)},\widetilde{Q}^{R,\mp}_{(1)} \bigr\}_{q{\rm P}} \bigr\}_{q{\rm P}}Q^{R,\pm}_{(1)}\,. \nonumber 
\end{eqnarray}
The fourth bracket is evaluated as 
\begin{eqnarray}
&&\bigl\{Q^{R,\pm}_{(1)},\bigl\{Q^{R,\pm}_{(1)},\bigl\{Q^{R,\pm}_{(1)},\widetilde{Q}^{R,\mp}_{(1)}\bigr\}_{q{\rm P}}\bigr\}_{q{\rm P}}\bigr\}_{q{\rm P}} \nonumber \\
&=&\bigl\{Q^{R,\pm}_{(1)},\bigl\{Q^{R,\pm}_{(1)},\bigl\{Q^{R,\pm}_{(1)},\widetilde{Q}^{R,\mp}_{(1)}\bigr\}_{q{\rm P}}\bigr\}_{q{\rm P}}\bigr\}_{\rm P}
+i\gamma\bigl\{Q^{R,\pm}_{(1)},\bigl\{Q^{R,\pm}_{(1)},\widetilde{Q}^{R,\mp}_{(1)}\bigr\}_{q{\rm P}}\bigr\}_{q{\rm P}}Q^{R,\pm}_{(1)} \nonumber \\
&=&\bigl\{Q^{R,\pm}_{(1)},\bigl\{Q^{R,\pm}_{(1)},\bigl\{Q^{R,\pm}_{(1)},\widetilde{Q}^{R,\mp}_{(1)}\bigr\}_{q{\rm P}}\bigr\}_{\rm P}\bigr\}_{\rm P}
+i\gamma\bigl\{Q^{R,\pm}_{(1)},\bigl\{Q^{R,\pm}_{(1)},\widetilde{Q}^{R,\mp}_{(1)}\bigr\}_{q{\rm P}}\bigr\}_{\rm P}Q^{R,\pm}_{(1)} \nonumber \\
&=&\bigl\{Q^{R,\pm}_{(1)},\bigl\{Q^{R,\pm}_{(1)},\bigl\{Q^{R,\pm}_{(1)},\widetilde{Q}^{R,\mp}_{(1)}\bigr\}_{\rm P}
-i\gamma\widetilde{Q}^{R,\mp}_{(1)}Q^{R,\pm}_{(1)}\bigr\}_{\rm P}\bigr\}_{\rm P} \nonumber \\
& & \quad 
+i\gamma\bigl\{Q^{R,\pm}_{(1)},\bigl\{Q^{R,\pm}_{(1)},\widetilde{Q}^{R,\mp}_{(1)}\bigr\}_{\rm P}
-i\gamma\widetilde{Q}^{R,\mp}_{(1)}Q^{R,\pm}_{(1)}\bigr\}_{\rm P}Q^{R,\pm}_{(1)} \nonumber \\
&=&\bigl\{Q^{R,\pm}_{(1)},\bigl\{Q^{R,\pm}_{(1)},\bigl\{Q^{R,\pm}_{(1)},\widetilde{Q}^{R,\mp}_{(1)}
\bigr\}_{\rm P}\bigr\}_{\rm P}\bigr\}_{\rm P} 
+ \gamma^2\bigl\{Q^{R,\pm}_{(1)},\widetilde{Q}^{R,\mp}_{(1)}\bigr\}_{\rm P}(Q^{R,\pm}_{(1)})^2  \nonumber\,.
\end{eqnarray}
With the Poisson brackets in (\ref{Poisson nonlocal}), one can show the following:
\begin{eqnarray}
&&\bigl\{Q^{R,\pm}_{(1)},\bigl\{Q^{R,\pm}_{(1)},\bigl\{Q^{R,\pm}_{(1)},\widetilde{Q}^{R,\mp}_{(1)}\bigr\}_{\rm P}\bigr\}_{\rm P}\bigr\}_{\rm P} 
+\gamma^2\bigl\{Q^{R,\pm}_{(1)},\widetilde{Q}^{R,\mp}_{(1)}\bigr\}_{\rm P}(Q^{R,\pm}_{(1)})^2  
= 0\,. \nonumber 
\end{eqnarray}
Thus the relation (\ref{q-serre1}) has been proven. 
Similarly, one can easily show the relation (\ref{q-serre2}).

\end{document}